\documentclass[reprint,amsmath,amssymb,superscriptaddress]{revtex4-1}

\usepackage{graphicx}
\usepackage{dcolumn}
\usepackage{bm}
\usepackage{color}
\begin{document}

%\preprint{}

\title{Giant interfacial perpendicular magnetic anisotropy in Fe/CuIn$_{1-x}$Ga$_x$Se$_2$\\ beyond Fe/MgO}


\author{Keisuke Masuda}
%\email{MASUDA.Keisuke@nims.go.jp}
\affiliation{Research Center for Magnetic and Spintronic Materials, National Institute for Materials Science (NIMS), 1-2-1 Sengen, Tsukuba 305-0047, Japan}

\author{Shinya Kasai}
\affiliation{Research Center for Magnetic and Spintronic Materials, National Institute for Materials Science (NIMS), 1-2-1 Sengen, Tsukuba 305-0047, Japan}

\author{Yoshio Miura}
%\email{miura@kit.ac.jp}
\affiliation{Research Center for Magnetic and Spintronic Materials, National Institute for Materials Science (NIMS), 1-2-1 Sengen, Tsukuba 305-0047, Japan}
\affiliation{Kyoto Institute of Technology, Electrical Engineering and Electronics, Kyoto 606-8585, Japan}
\affiliation{Center for Materials Research by Information Integration, National Institute for Materials Science (NIMS), 1-2-1 Sengen, Tsukuba 305-0047, Japan}
\affiliation{Center for Spintronics Research Network (CSRN), Graduate School of Engineering Science, Osaka University, Machikaneyama 1-3, Toyonaka, Osaka 560-8531, Japan}

\author{Kazuhiro Hono}
\affiliation{Research Center for Magnetic and Spintronic Materials, National Institute for Materials Science (NIMS), 1-2-1 Sengen, Tsukuba 305-0047, Japan}
%\thanks{}
%\affiliation{}


\date{\today}

\begin{abstract}
We study interfacial magnetocrystalline anisotropies in various Fe/semiconductor heterostructures by means of first-principles calculations. We find that many of those systems show perpendicular magnetic anisotropy (PMA) with a positive value of the interfacial anisotropy constant $K_{\rm i}$. In particular, the Fe/CuInSe$_2$ interface has a large $K_{\rm i}$ of $\sim 2.3\,{\rm mJ/m^2}$, which is about 1.6 times larger than that of Fe/MgO known as a typical system with relatively large PMA. We also find that the values of $K_{\rm i}$ in almost all the systems studied in this work follow the well-known Bruno's relation, which indicates that minority-spin states around the Fermi level provide dominant contributions to the interfacial magnetocrystalline anisotropies. Detailed analyses of the local density of states and wave-vector-resolved anisotropy energy clarify that the large $K_{\rm i}$ in Fe/CuInSe$_2$ is attributed to the preferable $3d$-orbital configurations around the Fermi level in the minority-spin states of the interfacial Fe atoms. Moreover, we have shown that the locations of interfacial Se atoms are the key for such orbital configurations of the interfacial Fe atoms.
\end{abstract}

\pacs{}
%\keywords{}

\maketitle

\section{\label{introduction} introduction}
Magnetic tunneling junctions (MTJs), in which an insulator barrier is sandwiched between two ferromagnetic electrodes, are the most important practical spintronic devices. They are currently used in non-volatile magnetic random access memories (MRAM), read heads of hard disk drives (HDD), and other magnetic sensors. For all these applications, high magnetoresistance (MR) ratios are required for high-voltage output as magnetic sensors. Second, low resistance-area products ($RA$) are essential for achieving high recording densities in HDD and MRAM. Moreover, we have an additional requirement for spin transfer torque MRAM (STT-MRAM) \cite{2016Dieny-Wiley} and voltage-controlled MRAM \cite{2016Shiota-APEX} applications that MTJs need to have perpendicular magnetic anisotropy (PMA) at interfaces between the electrode and the barrier layers. In particular, for STT-MRAM, PMA is indispensable to reduce the critical current for STT switching with sufficient thermal stability \cite{2016Dieny-Wiley}. Such MTJs with interfacial PMA \cite{2008Yoshikawa-IEEE,2009Mizunuma-APL,2010Yakushiji-APL,2017Yakushiji-APL} are referred to as p-MTJs.

In the early stages of research on interfacial magnetocrystalline anisotropy, magnetization measurements showed that the interfaces of Co/Pt and Co/Pd heterostructures exhibit PMA with an interfacial anisotropy constant $K_{\rm i}$ smaller than $1\,{\rm mJ/m^2}$ \cite{1985Carcia-APL,1987Draaisma-JMMM,1988Carcia-JAP}. It has also been clarified from x-ray magnetic circular dichroism (XMCD) spectroscopy that the PMA is attributed to the enhanced orbital moment of Co, which is induced by the hybridization between Co 3$d$ and Pt 5$d$ (Pd 4$d$) states \cite{1994Weller-PRB,1998Nakajima-PRL}. Thus, it has been believed that heavy elements with 5$d$ or 4$d$ electrons are required to obtain interfacial PMA.

In the past two decades, PMA has also been observed in heterostructures with oxide barriers \cite{2002Monso-APL,2003Rodmacq-JAP,2008Manchon-JAP,2009Nistor-APL,2013Koo-APL,2016Nozaki-PRApplied,2010Ikeda-NatMat}. Several studies have found perpendicular magnetic anisotropy at the interface of Co(Fe)/MO$_x$ ($M=$ Al, Mg, Cr, Ta, Ru, etc.) under appropriate oxidation conditions \cite{2002Monso-APL,2003Rodmacq-JAP,2008Manchon-JAP,2009Nistor-APL}. Moreover, large PMA with $K_{\rm i}>1\,{\rm mJ/m^2}$ has been achieved in Fe/MgO \cite{2013Koo-APL,2016Nozaki-PRApplied} and Fe-rich CoFeB/MgO \cite{2010Ikeda-NatMat}. These results suggest that the coupling between O and Co(Fe) atoms at interfaces of heterostructures plays a crucial role in PMA, which was actually confirmed by x-ray photoelectron spectroscopy (XPS) measurements \cite{2008Manchon-JAP}.

Recently, Kasai {\it et al.} \cite{2016Kasai-APL} succeeded in fabricating novel MTJs where semiconductor CuIn$_{0.8}$Ga$_{0.2}$Se$_{2}$ (CIGS) is sandwiched between ferromagnetic electrodes. They found that the CIGS-based MTJs have both high MR ratios ($100\%$ at $8\,{\rm K}$ and $40\%$ at room temperature) and low $RA$ values ($0.3$-$3\, \Omega\, {\mu {\rm m}}^2$) \cite{2016Kasai-APL}. Such high MR output was explained theoretically as a result of the spin-dependent coherent tunneling of $\Delta_{1}$ wave functions \cite{2017Masuda-JJAP}. In a more recent study, Mukaiyama {\it et al.} demonstrated large voltage outputs under bias voltages, which indicated that the CIGS-based MTJ is particularly attractive for read-head applications \cite{2017Mukaiyama-APEX}. To consider potential applications to STT-MRAM cells, the possibility of obtaining large PMA on CIGS-based MTJs must be investigated; however, no experimental and theoretical studies have been done on this issue.

In the present work, we investigate interfacial magnetocrystalline anisotropies of various Fe/semiconductor heterostructures including Fe/CuIn$_{1-x}$Ga$_{x}$Se$_2$ by means of first-principles calculations. We found that all of the Fe/CuIn$_{1-x}$Ga$_{x}$Se$_2$ heterostructures show PMA. In particular, the Fe/CuInSe$_2$ system exhibits a quite large $K_{\rm i} \approx 2.3\,{\rm mJ/m^2}$, which is approximately 1.6 times as large as that of the Fe/MgO system that is currently used for p-MTJs. We also found that Bruno's relation \cite{1989Bruno-PRB}, which states that magnetocrystalline anisotropy is proportional to the anisotropy in the orbital magnetization of a ferromagnet, holds for almost all the systems considered in this study. This suggests that magnetocrystalline anisotropies can be described by second-order perturbation theory with respect to spin-orbit interactions, in which electron scatterings only in minority-spin states are considered. By analyzing the local density of states (LDOS) and wave-vector-resolved interfacial anisotropy, we show that the large $K_{\rm i}$ in Fe/CuInSe$_2$ can be understood naturally from the orbital configurations in the minority-spin states near the Fermi level.

\section{\label{methods} calculation method}
\begin{figure}
\includegraphics[width=8.3cm]{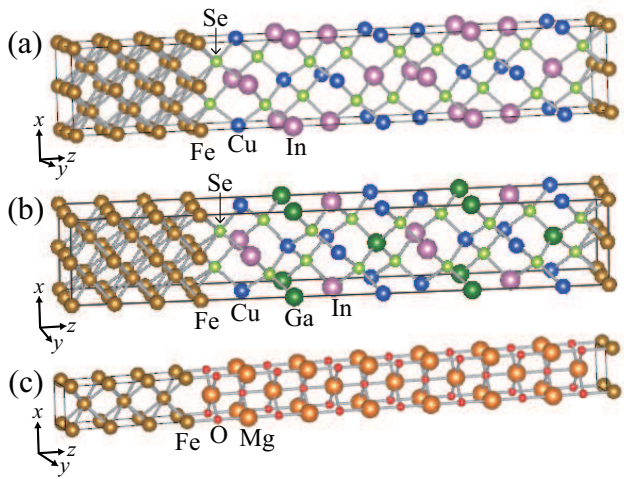}
\caption{\label{supercells} Supercells of (a) Fe(7)/CuInSe$_2$(17), (b) Fe(7)/CuIn$_{0.5}$Ga$_{0.5}$Se$_2$(17), and (c) Fe(7)/MgO(13).}
\end{figure}
We carried out first-principles calculations on the basis of density-functional theory (DFT) including spin-orbit interactions, which is implemented in the Vienna $ab$ $initio$ simulation program (VASP) \cite{1996Kresse-PRB}. For the exchange-correlation energy, we adopted the spin-polarized generalized gradient approximation (GGA) proposed by Perdew, Becke, and Ernzerhof \cite{1996Perdew-PRL}. The projector augmented wave (PAW) potential \cite{1994Bloechl-PRB,1999Kresse-PRB} was also used to take into account the effect of core electrons properly. In this work, we considered 14 types of Fe/semiconductor(001) heterostructures, the list of which is given in Table \ref{tab1}. We also considered an Fe/MgO(001) heterostructure to obtain a benchmark of $K_{\rm i}$ under the present calculation conditions. First, we prepared a supercell Fe(7)/$X$(17) for each heterostructure with semiconductor $X$ as shown in Figs. \ref{supercells}(a) and \ref{supercells}(b), where each number in the parentheses represents the number of layers. For the Fe/MgO(001) heterostructure, we used a supercell Fe(7)/MgO(13) [see Fig. \ref{supercells}(c)] because the thickness of MgO(13) is close to those of the semiconductors $X$(17). Since the barrier is sufficiently thicker than the Fe electrode in all the supercells, the in-plane lattice constant $a$ of each supercell was fixed to the experimental lattice constant of the barrier $a_{\rm barrier}$ shown in Table \ref{tab1}. Note here that we can set $a=a_{\rm barrier}/\sqrt{2}$ in the cases of ZnSe, ZnS, GaAs, and MgO barriers due to the high symmetry of the structures. As experimental lattice constants of ternary chalcopyrite semiconductors, we adopted the values in Ref. \cite{1984Jaffe-PRB}. We also used the equation $a_{{\rm CuIn}_{1-x}{\rm Ga}_{x}{\rm Se}_{2}}=(1-x) \times a_{\rm CuInSe_{2}}+x \times a_{\rm CuGaSe_{2}}$ to set the experimental lattice constant of CuIn$_{1-x}$Ga$_{x}$Se$_2$. Unfortunately, VASP cannot treat disorder between In and Ga atoms in CuIn$_{1-x}$Ga$_x$Se$_2$. Thus, we treated this as a percentage of the numbers of atoms in the supercell. For example, in the case of Fe/CuIn$_{0.5}$Ga$_{0.5}$Se$_2$(001), we assigned the same number of atomic sites for In and Ga in the supercell, where the atomic configurations of these atoms were chosen as shown in Fig. \ref{supercells}(b). In all the supercells, we optimized each atomic position and the distance between the barrier and the Fe electrode so that the total energy of the supercell is minimized. Such optimizations reduced the energy of each supercell by $1 \sim 3\,{\rm eV}$. In addition, we confirmed that Se, S, or As layers are energetically favored as the interfacial layers in all the Fe/semiconductor(001) heterostructures as shown in Figs. \ref{supercells}(a) and \ref{supercells}(b) (see Appendix \ref{terminationlayers} for details). The interfacial anisotropy constant $K_{\rm i}$ was calculated using the force theorem as $K_{\rm i}=(E_{[100]}-E_{[001]})/2S$, where $E_{[100]}$ ($E_{[001]}$) is the total energy of the supercell for the magnetization along the [100] ([001]) direction, $S$ is the cross-sectional area of the supercell, and the factor 2 in the denominator reflects the fact that two interfaces are included in the supercell. In this definition, a positive (negative) $K_{\rm i}$ shows a tendency toward perpendicular (in-plane) magnetic anisotropy. The list of ${\bf k}$ points used in the calculations of $K_{\rm i}$ is provided in Table \ref{tab1}. We used 10$\times$10$\times$1 ${\bf k}$ points for the heterostructures with ternary or quaternary chalcopyrite semiconductors. We used more ${\bf k}$ points for other heterostructures due to their smaller supercell sizes.
\begin{table}[t]
\caption{\label{tab1}
List of in-plane lattice constants $a$ of the heterostructures, ${\bf k}$ points used in the calculations of $K_{\rm i}$, and the obtained values of $K_{\rm i}$.
}
\begin{ruledtabular}
\begin{tabular}{lccc}
\textrm{$X$ in Fe/$X$(001)}&
\textrm{$a$\,(\AA)}&
\textrm{{\bf k} points}&
\textrm{$K_{\rm i}$\,(${\rm mJ/m^2}$)}\\
\colrule
ZnSe & 4.013 & 15$\times$15$\times$1 & 1.701\\
ZnS & 3.823 & 15$\times$15$\times$1 & 1.151\\
GaAs & 3.998 & 15$\times$15$\times$1 & 0.210\\
CuInSe$_2$ & 5.782 & 10$\times$10$\times$1 & 2.305\\
CuIn$_{0.75}$Ga$_{0.25}$Se$_2$ & 5.726 & 10$\times$10$\times$1 & 2.069\\
CuIn$_{0.5}$Ga$_{0.5}$Se$_2$ & 5.698 & 10$\times$10$\times$1 & 1.712\\
CuIn$_{0.25}$Ga$_{0.75}$Se$_2$ & 5.670 & 10$\times$10$\times$1 & 1.575\\
CuGaSe$_2$ & 5.614 & 10$\times$10$\times$1 & 1.266\\
CuInS$_2$ & 5.523 & 10$\times$10$\times$1 & -0.373\\
CuGaS$_2$ & 5.356 & 10$\times$10$\times$1 & 0.776\\
AgInSe$_2$ & 6.109 & 10$\times$10$\times$1 & 0.721\\
AgGaSe$_2$ & 5.985 & 10$\times$10$\times$1 & 1.257\\
AgInS$_2$ & 5.872 & 10$\times$10$\times$1 & 0.841\\
AgGaS$_2$ & 5.754 & 10$\times$10$\times$1 & 0.027\\
MgO & 2.982 & 20$\times$20$\times$1 & 1.396\\
\end{tabular}
\end{ruledtabular}
\end{table}

\section{\label{resultsdiscussion} results and discussion}
\begin{figure}
\includegraphics[width=7.8cm]{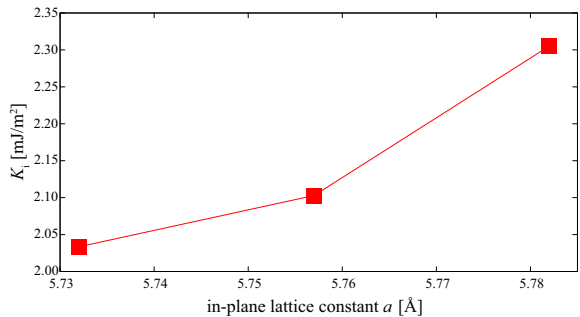}
\caption{\label{a_vs_K} In-plane lattice constant dependence of the interfacial anisotropy constant $K_{\rm i}$ for Fe/CuInSe$_2$(001).}
\end{figure}
\begin{figure}
\includegraphics[width=8.7cm]{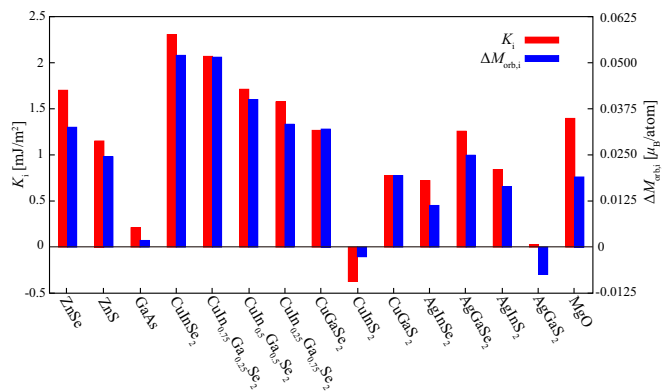}
\caption{\label{K_dMo} Correlation between the interfacial anisotropy constant $K_{\rm i}$ and anisotropy of the orbital magnetic moment at the interfacial Fe layer $\Delta M_{\rm orb,i}$ (see text for details).}
\end{figure}
\begin{figure}
\includegraphics[width=4.5cm]{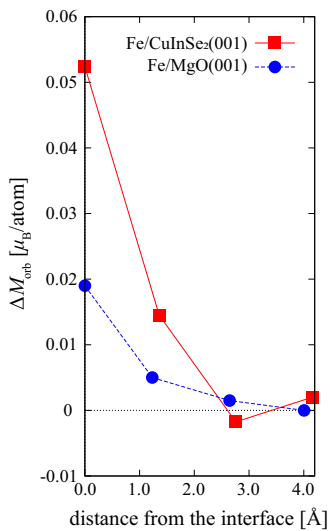}
\caption{\label{dMo_n-Fe} Anisotropy of the orbital magnetic moment $\Delta M_{\rm orb}$ resolved into each Fe-layer contribution for Fe/CuInSe$_2$(001) and Fe/MgO(001). The horizontal axis shows the distance of each Fe layer from the interface.}
\end{figure}
We show the obtained values of $K_{\rm i}$ in Table \ref{tab1}. We find that all systems except Fe/CuInS$_2$(001) have positive values of $K_{\rm i}$. In particular, Fe/CuInSe$_2$(001) has the largest $K_{\rm i}$ of $2.305\,{\rm mJ/m^2}$, which is about 1.6 times as large as our benchmark value $1.396\,{\rm mJ/m^2}$ in Fe/MgO(001). Similar to our previous study \cite{2017Masuda-JJAP}, we additionally considered the effect of the Coulomb interaction $U$ in the Cu $3d$ states of Fe/CuInSe$_2$(001) on $K_{\rm i}$. We obtained $K_{\rm i}=2.363$ and $2.403\,{\rm mJ/m^2}$ for $U=5$ and $10\,{\rm eV}$, respectively, which indicates that the interaction $U$ in the barriers does not have significant effects on $K_{\rm i}$. We also studied the in-plane lattice constant $a$ dependence of $K_{\rm i}$ for Fe/CuInSe$_2$(001), as shown in Fig. \ref{a_vs_K}. The possible smallest value of $a$ is considered to be twice the lattice constant of bulk bcc Fe, $a=2a_{\rm Fe}=5.732\,{\rm \AA}$. Thus, we changed $a$ from $2a_{\rm Fe}$ to $a_{\rm CuInSe_2}$. Note that this range includes the value of $a$ that is compatible with bcc Cr often used as buffer layers, $a=2a_{\rm Cr}=5.768\,{\rm \AA}$. We see that $K_{\rm i}$ changes smoothly with the value of $a$ and is over $2\,{\rm mJ/m^2}$ for all values of $a$ in the considered range.

As mentioned in Sec. \ref{methods}, a positive $K_{\rm i}$ indicates a tendency toward PMA. However, to be more precise, the following effective anisotropy should be used for a more accurate estimation of PMA: $K_{\rm eff}t_{\rm eff}=K_{\rm i}-2 \pi M^2_{\rm s} t_{\rm eff}$, where $M_{\rm s}$ is the saturation magnetization and $t_{\rm eff}$ is the effective thickness of a ferromagnetic electrode. The second term $2 \pi M^2_{\rm s} t_{\rm eff}$ is the contribution from magnetic shape anisotropy, which always favors in-plane magnetization. In our present situation, since bcc Fe has a magnetization of $2.262\, \mu_{\rm B}$ per atom and $t_{\rm eff}=t_{\rm Fe}/2 \approx 0.425\,{\rm nm}$, the shape anisotropy term is estimated as $2 \pi M^2_{\rm s} t_{\rm eff} \sim 0.85\,{\rm mJ/m^2}$, which does not exceed $K_{\rm i}$ in many systems considered in this study \cite{remark1}. Therefore, we can conclude that many Fe/semiconductor(001) heterostructures favor PMA even if we use $K_{\rm eff}t_{\rm eff}$ for an estimation of interfacial magnetocrystalline anisotropy.

In Fig. \ref{K_dMo}, we show the correlation between the interfacial anisotropy constant $K_{\rm i}$ and the anisotropy of the orbital magnetic moment in the interfacial Fe atom $\Delta M_{\rm orb,i}$ for heterostructures studied in this work. Here, the anisotropy of the orbital magnetic moment is defined by $\Delta M_{\rm orb,i}=M^{[001]}_{\rm orb,i}-M^{[100]}_{\rm orb,i}$, where $M^{[001]}_{\rm orb,i}$ ($M^{[100]}_{\rm orb,i}$) is the orbital magnetic moment of the interfacial Fe atom for magnetization along the [001] ([100]) direction. We clearly see that the so-called Bruno's relation $K_{\rm i} \propto \Delta M_{\rm orb,i}$ \cite{1989Bruno-PRB} holds for almost all systems considered in this study. A previous theoretical work has confirmed this relation in the Fe(Co)/MgO interface \cite{2014Zhang-PRB}. Note, however, that values of $K_{\rm i}$ and $\Delta M_{\rm orb,i}$ in Fe-based heterostructures do not always follow the Bruno's relation, as shown in a systematic theoretical study \cite{2013Miura-JAP}. Laan \cite{1998Laan-JPCM} indicated that the Bruno's relation is satisfied when no spin-flip scattering occurs and majority-spin states are fully occupied. In such a case, the minority-spin scattering between unoccupied and occupied states around the Fermi level ($E_{\rm F}$) provides the dominant contribution to the magnetocrystalline anisotropy \cite{1993Wang-PRB}. As shown later, Fe/CuInSe$_2$(001) has a suitable LDOS and band structure to yield large PMA through such minority-spin scattering. Figure \ref{dMo_n-Fe} shows the anisotropy of the orbital magnetic moment $\Delta M_{\rm orb}$ resolved into each Fe-layer contribution for Fe/CuInSe$_2$(001) and Fe/MgO(001). In both systems, the interfacial Fe layer has a much larger $\Delta M_{\rm orb}$ compared to other layers. Since the magnetic anisotropy is proportional to $\Delta M_{\rm orb}$ in these systems, as mentioned above, the results in Fig. \ref{dMo_n-Fe} clearly indicate that the anisotropy $K_{\rm i}$ is mainly due to the interfacial contribution.

\begin{figure}
\includegraphics[width=8.5cm]{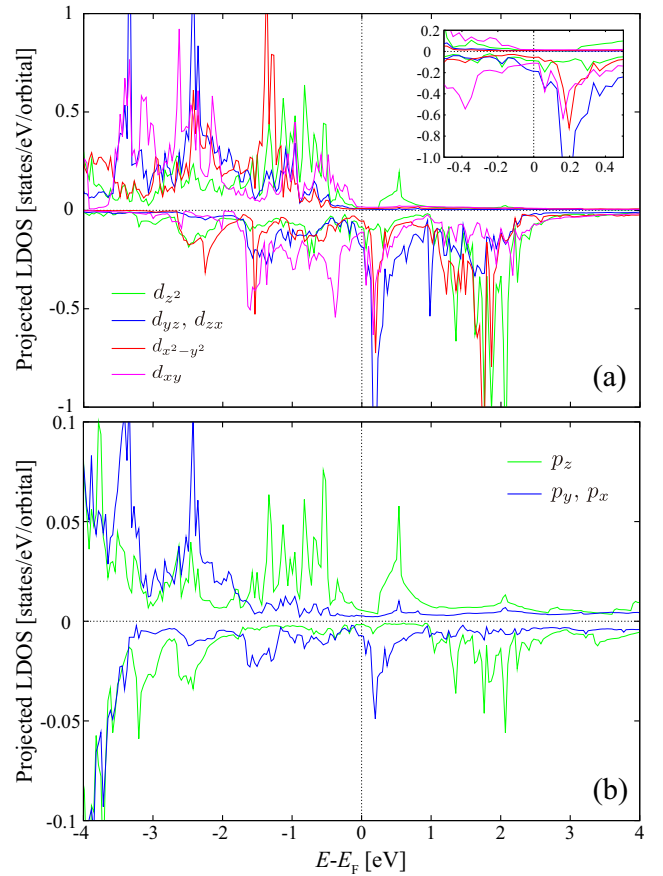}
\caption{\label{MgO_ldos} The projected LDOSs for (a) Fe 3$d$ states and (b) O 2$p$ states at the interface of the Fe/MgO(001) heterostructure. In each panel, positive and negative values indicate the majority- and minority-spin projected LDOSs, respectively. The inset of panel (a) shows an enlarged view near the Fermi level.}
\end{figure}
\begin{figure}
\includegraphics[width=8.5cm]{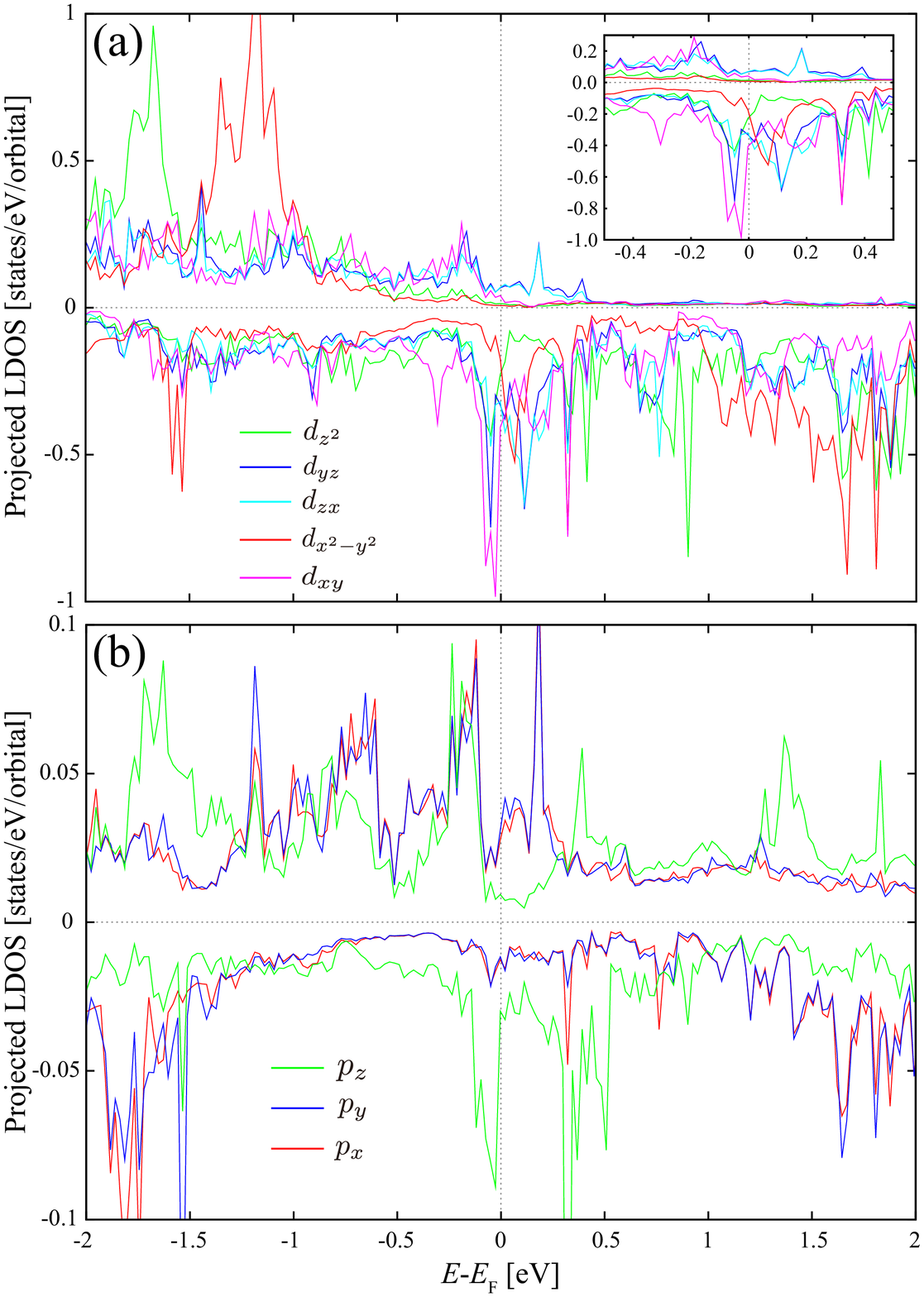}
\caption{\label{CIS_ldos} The projected LDOSs for (a) Fe 3$d$ states and (b) Se 4$p$ states at the interface of the Fe/CuInSe$_{2}$(001) heterostructure. In each panel, positive and negative values indicate the majority- and minority-spin projected LDOSs, respectively. The inset of panel (a) shows an enlarged view near the Fermi level.}
\end{figure}
\begin{figure}
\includegraphics[width=8.5cm]{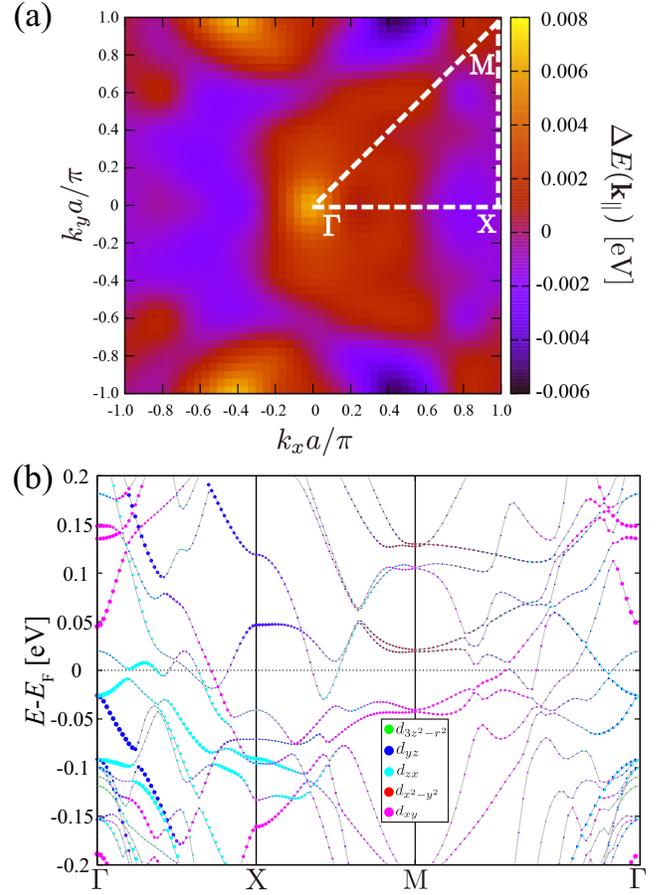}
\caption{\label{k-dE_banddn} The direct microscopic information on PMA in the Fe/CuInSe$_{2}$(001) heterostructure: (a) the in-plane wave vector (${\bf k}_{\parallel}$) dependence of $\Delta E({\bf k}_{\parallel}) \equiv E_{[100]}({\bf k}_{\parallel})-E_{[001]}({\bf k}_{\parallel})$ and (b) the minority-spin bands along the high symmetry lines in the ${\bf k}_{\parallel}$ Brillouin zone. In panel (b), orbital components of each band are indicated by colors.}
\end{figure}
\begin{figure}[t]
\includegraphics[width=8.5cm]{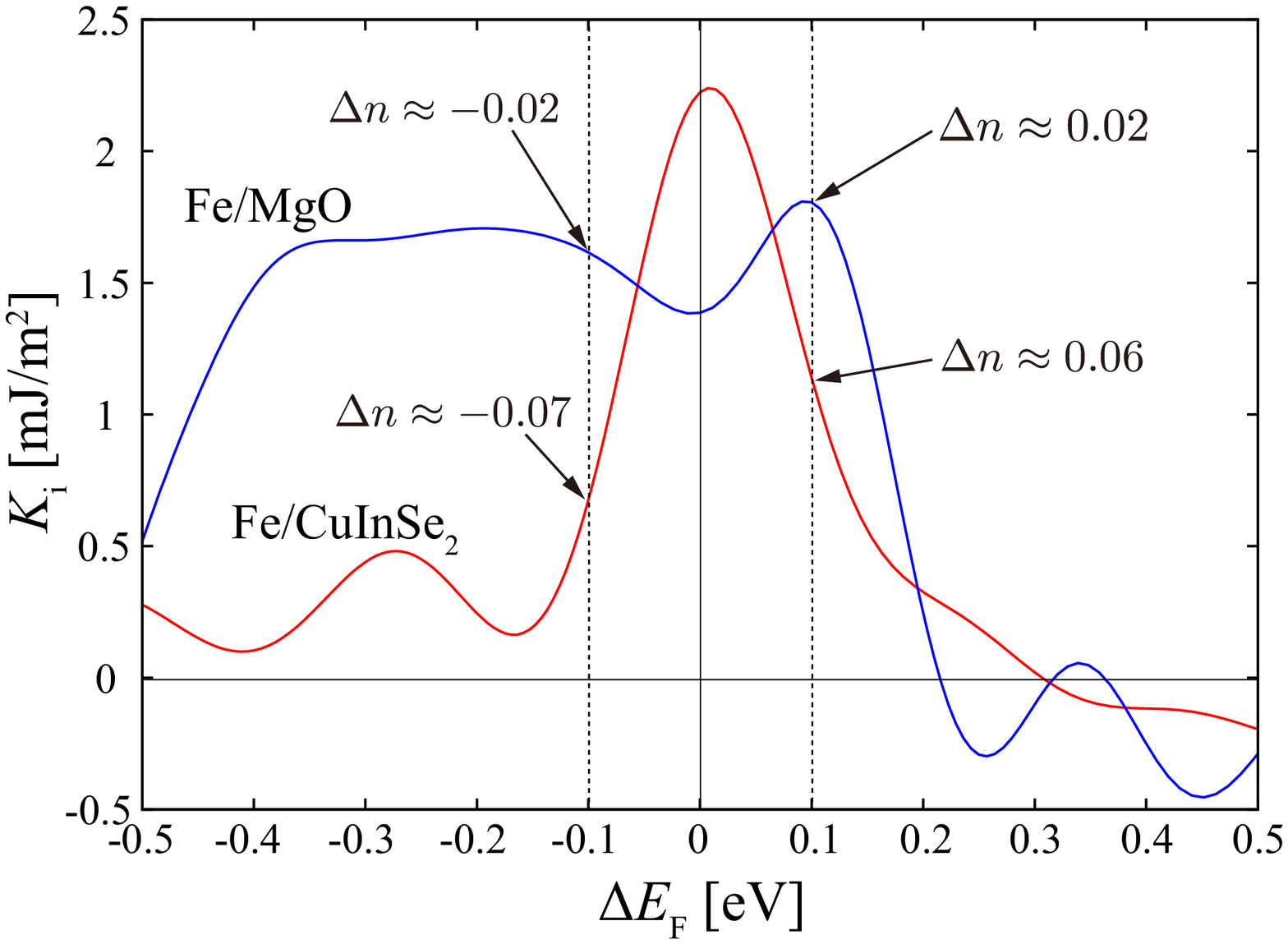}
\caption{\label{CIS-MgO_dEFvsK} Calculated $K_{\rm i}$ as a function of the change in the Fermi energy $\Delta E_{\rm F}$ for Fe/CuInSe$_2$(001) and Fe/MgO(001). The origin of the horizontal axis $\Delta E_{\rm F}=0$ corresponds to the original Fermi energy in each system. The changes in the valence electron number $\Delta n$ at $\Delta E_{\rm F}=\pm 1\,{\rm eV}$ are also shown in units of electrons/atom.}
\end{figure}
Before discussing Fe/CuInSe$_2$(001) with the largest $K_{\rm i}$, let us first focus on the Fe/MgO(001) for comparison. Previous theoretical studies \cite{2010Nakamura-PRB,2010Niranjan-APL,2013Hallal-PRB,2011Yang-PRB} have shown that this system has relatively large PMA with $K_{\rm i}=1 \sim 2\, {\rm mJ/m^2}$, which is consistent with our present results $K_{\rm i}=1.396\, {\rm mJ/m^2}$. Figure \ref{MgO_ldos}(a) shows the calculated LDOS for $3d$ states of an Fe atom located at the interface between Fe and MgO layers, whose main features around $E_{\rm F}$ are consistent with previous results on similar systems \cite{2014Yoshikawa-APEX, 2013Khoo-PRB}. To understand the relationship between the LDOS and the magnetic anisotropy constant $K_{\rm i}$, we introduce the following expression for $K_{\rm i}$ derived from the second-order perturbation expansion with respect to the spin-orbit interaction \cite{1993Wang-PRB}:
\begin{equation}
K_{\rm i} \approx \xi^2 \sum_{o_{\downarrow},u_{\downarrow}} \frac{|\langle o_{\downarrow} |L_{z}| u_{\downarrow} \rangle |^{2} - |\langle o_{\downarrow} |L_{x}| u_{\downarrow} \rangle |^{2}}{\epsilon_{u_{\downarrow}}-\epsilon_{o_{\downarrow}}} ,\label{eq:K}
\end{equation}
where $\xi$ is the coupling constant of the LS spin-orbit interaction, $|o_{\downarrow}\rangle$ ($|u_{\downarrow}\rangle$) is an occupied (unoccupied) state with minority spin, $\epsilon_{o_{\downarrow}}$ ($\epsilon_{u_{\downarrow}}$) is the energy of the $|o_{\downarrow}\rangle$ ($|u_{\downarrow}\rangle$) state, and $L_{\alpha}$ ($\alpha=x,z$) are the usual angular momentum operators. In Eq. (\ref{eq:K}), we considered excitation processes between the minority-spin occupied and unoccupied states, and we neglected small contributions from spin-flip scattering processes. This approach was shown to be sufficient to understand PMA in Fe/MgO(001) systems \cite{2010Nakamura-PRB}. We can easily see that the matrix element of $L_z$ ($L_x$) provides a positive (negative) contribution to $K_{\rm i}$. In addition, the excitation energy $\Delta \epsilon \equiv \epsilon_{u_{\downarrow}}-\epsilon_{o_{\downarrow}}$ is also an important factor: the excitation process with smaller $\Delta \epsilon$ contributes more significantly to $K_{\rm i}$. In the minority-spin LDOS of an interfacial Fe atom in Fig. \ref{MgO_ldos}(a), the $d_{xy}$ state has a peak a little below $E_{\rm F}$ ($E-E_{\rm F} \approx -0.4\,{\rm eV}$), and the $d_{yz}$, $d_{zx}$, and $d_{x^2-y^2}$ states have peaks just above $E_{\rm F}$ ($E-E_{\rm F} \approx 0.2\,{\rm eV}$). These states yield finite values of $\langle d_{xy} | L_{z} | d_{x^2-y^2} \rangle$ and $\langle d_{xy} | L_{x} | d_{yz} (d_{zx}) \rangle$. As a result, positive contributions from $\langle L_z \rangle$ exceed negative ones from $\langle L_x \rangle$, leading to PMA with positive $K_{\rm i}$. Figure \ref{MgO_ldos}(b) shows the LDOSs of $2p$ states in an interfacial O atom. Note that finite values of LDOSs occur around $E_{\rm F}$ although bulk MgO is a band insulator. Such states are induced by metallic states of interfacial Fe atoms and are thus called metal-induced gap states (MIGS). The concept of MIGS was first introduced as a way to understanding metal/semiconductor interfaces \cite{1965Heine-PR,1976Louie-PRB}, and it was later applied to Cu/MgO \cite{1998Muller-PRL} and Fe/MgO \cite{2001Butler-PRB} interfaces. By comparing Figs. \ref{MgO_ldos}(a) and \ref{MgO_ldos}(b), we see that structures of LDOSs around $E_{\rm F}$ are quite similar between Fe $d_{3z^{2}-r^2}$ and O $p_z$ states, which is due to the strong hybridization between these states at the interface. Such strong hybridization comes from the geometry of the interface, where O atoms are on top of Fe atoms in the $z$ direction, as seen in Fig. \ref{supercells}(c). We also see that in the minority-spin LDOSs, the peak structure of the Fe $d_{yz} (d_{zx})$ state just above $E_{\rm F}$ is almost the same as that of the O $p_{y} (p_x)$ state.

Let us now discuss the relationship between the large positive $K_{\rm i}$ and LDOSs in Fe/CuInSe$_2$(001). Figure \ref{CIS_ldos}(a) shows the LDOSs for 3$d$ states of an Fe atom located at the interface between Fe and CuInSe$_2$ layers. We can readily identify some sharp peaks in minority-spin LDOSs both just above and just below $E_{\rm F}$. Such LDOSs structures enable excitations with quite small $\Delta \epsilon$, which provide large contributions to $K_{\rm i}$ as mentioned above. As seen from the orbital configurations around $E_{\rm F}$ shown in the inset of Fig. \ref{CIS_ldos}(a), these peaks can yield finite values of $\langle d_{xy} | L_{z} | d_{x^2-y^2} \rangle$, $\langle d_{zx} | L_{z} | d_{yz} \rangle$, $\langle d_{yz} | L_{z} | d_{zx} \rangle$, and $\langle d_{xy} | L_{x} | d_{yz} (d_{zx}) \rangle$. Thus, Fe/CuInSe$_2$(001) has more excitation processes with positive contributions to $K_{\rm i}$ than Fe/MgO(001). Moreover, such positive processes have smaller $\Delta \epsilon$ than those of Fe/MgO(001). These features in the LDOSs support our results that $K_{\rm i}$ of Fe/CuInSe$_2$(001) is about 1.6 times larger than that of Fe/MgO(001). We emphasize that such preferable LDOSs of interfacial Fe have a close relationship with the LDOSs of interfacial Se shown in Fig. \ref{CIS_ldos}(b). Similar to the interfacial O LDOSs in Fe/MgO(001), Se LDOSs have finite values around $E_{\rm F}$ due to the MIGS. In the minority-spin Se LDOSs, we can find some sharp peaks around $E_{\rm F}$ with $p_z$-orbital character, which couple to various minority-spin Fe $3d$ states around $E_{\rm F}$ [see the inset of Fig. \ref{CIS_ldos}(a)]. This is in contrast to the Fe/MgO(001) case, in which $p_z$ states in interfacial O atoms couple mainly to $d_{3z^{2}-r^2}$ states in interfacial Fe atoms. Such a difference mainly comes from the difference in the geometry at the interface between Fe and barrier layers. In Fe/CuInSe$_2$(001), interfacial Se atoms are not on top of Fe atoms in the $z$ direction, unlike the Fe/MgO(001) case, as shown in Fig. \ref{supercells}(a). Owing to such locations of interfacial Se atoms, $p_z$ wave functions of interfacial Se can hybridize with almost all 3$d$ states of interfacial Fe, not only with $d_{3z^{2}-r^2}$ states; this yields favorable orbital configurations around $E_{\rm F}$.

We carried out further analysis to obtain more detailed information on large $K_{\rm i}$ in Fe/CuInSe$_2$(001). Figure \ref{k-dE_banddn}(a) shows the in-plane wave vector (${\bf k}_{\parallel}$) dependence of $\Delta E({\bf k}_{\parallel}) \equiv E_{[100]}({\bf k}_{\parallel})-E_{[001]}({\bf k}_{\parallel})$. Here, $K_{\rm i}$ is proportional to the sum of $\Delta E({\bf k}_{\parallel})$ over all ${\bf k}_{\parallel}$ in the Brillouin zone. We obtained positive values of $\Delta E({\bf k}_{\parallel})$ in wide regions of the Brillouin zone, including high-symmetry $\Gamma$ and M points. To understand the origin of the positive $\Delta E({\bf k}_{\parallel})$, we plotted in Fig. \ref{k-dE_banddn}(b) the band structure of the Fe/CuInSe$_2$(001) supercell along the high-symmetry lines. Around the $\Gamma$ point, both the highest occupied and lowest unoccupied bands  are generated by the hybridization between $d_{yz}$ and $d_{zx}$ states of Fe, which can enhance $\langle d_{zx} | L_{z} | d_{yz} \rangle$ and $\langle d_{yz} | L_{z} | d_{zx} \rangle$, giving positive contributions to $K_{\rm i}$. Around the M point, the highest occupied and lowest unoccupied bands consist of $d_{xy}$ and $d_{x^2-y^2}$ states of Fe, respectively. These bands can enhance $\langle d_{xy} | L_{z} | d_{x^2-y^2} \rangle$ with positive contributions to $K_{\rm i}$. On the other hand, we have small negative values of $\Delta E({\bf k}_{\parallel})$ around X point. As shown in Fig. \ref{k-dE_banddn}(b), since the highest occupied (lowest unoccupied) band around X point comes from $d_{zx}$ and $d_{xy}$ ($d_{yz}$ and $d_{xy}$) states, $\langle d_{zx} | L_{x} | d_{xy} \rangle$ and $\langle d_{xy} | L_{x} | d_{yz} \rangle$ contribute to such negative values of $\Delta E({\bf k}_{\parallel}\approx{\rm X})$. The smallness of $\Delta E({\bf k}_{\parallel}\approx{\rm X})$ is due to the relatively large energy difference between the highest occupied and lowest unoccupied bands [see the denominator of Eq. (\ref{eq:K})]. All these band structures around $E_{\rm F}$ are consistent with the orbital configurations in Fe LDOSs shown in the inset of Fig. \ref{CIS_ldos}(a).

Finally, we show in Fig. \ref{CIS-MgO_dEFvsK} the calculated $K_{\rm i}$ as a function of the change in the Fermi energy $\Delta E_{\rm F}$ for Fe/CuInSe$_2$(001) and Fe/MgO(001). These curves were obtained by changing the valence electron number $n$ in each system. We see that $K_{\rm i}$ in Fe/MgO(001) increases slightly for small hole and electron dopings. On the other hand, $K_{\rm i}$ in Fe/CuInSe$_2$(001) decreases a great deal for both hole- and electron-doped cases, which is consistent with our findings that the large $K_{\rm i}$ in this system is closely related to the sharp peaks in the LDOS around $E_{\rm F}$ in Fe minority-spin states [see Fig. \ref{CIS_ldos}(a)]. In combination with large $K_{\rm i}$, the sharp decrease in $K_{\rm i}$ by dopings in Fe/CuInSe$_2$(001) is useful for the voltage-assisted MRAM applications \cite{2016Dieny-Wiley,2016Shiota-APEX}.

\section{summary}
We investigated interfacial magnetocrystalline anisotropies of various heterostructures consisting of Fe and non-oxide semiconductor layers by using first-principles calculations. We found that most of those systems show PMA with a positive interfacial anisotropy constant $K_{\rm i}$. In particular, Fe/CuInSe$_2$ was found to have the largest $K_{\rm i} \approx 2.3\, {\rm mJ/m^2}$, which is approximately 1.6 times as large as that of Fe/MgO, being a benchmark system currently used in p-MTJs. We also found that the Bruno's relation holds for almost all systems considered in this study, which means that the interfacial magnetocrystalline anisotropies are determined mainly by minority-spin states around $E_{\rm F}$. By analyzing the LDOS and wave-vector-resolved anisotropy energy, we clarified that the large $K_{\rm i}$ in Fe/CuInSe$_2$ is due to the preferable $3d$-orbital configurations around $E_{\rm F}$ in the minority-spin states of interfacial Fe atoms. Moreover, we found that the positions of interfacial Se atoms play a key role in the appearance of such orbital configurations of interfacial Fe atoms.

{\it Note added in proof.} In Ref. \cite{remark1}, we commented on magnetic shape anisotropy estimated from the magnetostatic dipole-dipole interaction. After our manuscript was accepted, we found a minor error in our program for calculating the magnetic shape anisotropy. By using the revised program, we obtained the shape anisotropy of around 1.1 mJ/m$^2$ for the present Fe/semiconductor(001) superlattices, which is sufficiently lower than the crystalline magnetic anisotropy in many of these systems.

\appendix*

\section{\label{terminationlayers} INTERFACIAL LAYERS OF SUPERCELLS}
To determine the energetically favored interfacial layer of the supercell, we compared formation energies of the supercell for different termination layers of the semiconductor barrier. Here, we show the details of the procedure in the case of Fe/CuInSe$_2$(001). The relative stability between Se- and CuIn-terminated interfaces is estimated by using the following formation energy \cite{2008Miura-PRB,2011Miura-PRB}:
\begin{equation}
E^{\rm term}_{\rm form}=E^{\rm term}_{\rm tot}-\sum_{i} N_i \mu_i,\label{formene}
\end{equation}
where $E^{\rm term}_{\rm tot}$ is the total energy of the optimized supercell for each termination, $N_i$ is the number of atoms of the element $i$, and $\mu_i$ is its chemical potential. Note here that the chemical potential of each atom does not exceed the corresponding one in the bulk phase, i.e., the upper limit of $\mu_i$ is given by the bulk total energy per atom. In the present study, we used $\mu_{\rm Cu(bulk)}$, $\mu_{\rm In(bulk)}$, and $\mu_{\rm Se(bulk)}$ derived from bulk energies of fcc Cu, tetragonal In, and hexagonal Se, respectively. From the thermodynamic equilibrium condition $\mu_{\rm CuInSe_{2}}=\mu_{\rm Cu}+\mu_{\rm In}+2\mu_{\rm Se}$, we can obtain the lower limit of $\mu_{\rm Se}$ as $\mu_{\rm Se} \geq (\mu_{\rm CuInSe_2}-\mu_{\rm Cu(bulk)}-\mu_{\rm In(bulk)})/2$. Here, we adopt $\mu_{\rm CuInSe_{2}}$ derived from the bulk structure. We estimated the difference in the formation energy $E^{\rm diff}_{\rm form}=(E^{\rm Se-term}-E^{\rm CuIn-term})/2$ in the thermodynamically allowed range $(\mu_{\rm CuInSe_2}-\mu_{\rm Cu(bulk)}-\mu_{\rm In(bulk)})/2 \leq \mu_{\rm Se} \leq \mu_{\rm Se(bulk)}$, where the upper (lower) limit of $\mu_{\rm Se}$ corresponds to the Se-rich (Se-poor) case. As a result of the calculations, we obtained $E^{\rm diff}_{\rm form}=-2.43\,{\rm eV}$ and $-0.77\,{\rm eV}$ for the Se-rich and Se-poor cases, respectively. These results indicate that the Se-terminated interface is energetically preferred in Fe/CuInSe$_2$(001). We carried out similar analysis in other Fe/semiconductor(001) heterostructures to determine their accurate interfacial structures. As a result, it was found that Se, S, or As layers are energetically favored as interfacial layers in all of the present Fe/semiconductor(001) heterostructures.

\begin{acknowledgments}
The authors are grateful to H. Sukegawa and S. Mitani for useful discussions and critical comments. This work was partly supported by Grant-in-Aids for Scientific Research (S) (Grant No. 16H06332) and (B) (Grant No. 16H03852) from the Ministry of Education, Culture, Sports, Science and Technology, Japan, by NIMS MI2I, and also by the ImPACT Program of Council for Science, Technology and Innovation, Japan. The crystal structures of the supercells were visualized using VESTA \cite{2011Momma_JAC}.
\end{acknowledgments}


\end{document}